\documentclass[conference,compsoc]{IEEEtran}
\IEEEoverridecommandlockouts

\usepackage{color}
\usepackage{booktabs}

\usepackage{xcolor}
\usepackage{amsmath}
\usepackage[numbers,sort]{natbib}
\usepackage{enumitem}

\usepackage[T1]{fontenc}
\usepackage{marvosym}

\usepackage{xurl}

\def\BibTeX{{\rm B\kern-.05em{\sc i\kern-.025em b}\kern-.08em
    T\kern-.1667em\lower.7ex\hbox{E}\kern-.125emX}}

\newcommand{\ignore}[1]{}

\author{
\IEEEauthorblockN{Mengying Wu\IEEEauthorrefmark{2}, Pei Chen\IEEEauthorrefmark{2}, Geng Hong\IEEEauthorrefmark{2}\Letter, Baichao An\IEEEauthorrefmark{2}, Jinsong Chen\IEEEauthorrefmark{2}, Binwang Wan\IEEEauthorrefmark{2}, Xudong Pan\IEEEauthorrefmark{2}\IEEEauthorrefmark{3}, }
\IEEEauthorblockN{
Jiarun Dai\IEEEauthorrefmark{2}, and Min Yang\IEEEauthorrefmark{2}\Letter}
\IEEEauthorblockA{
\IEEEauthorrefmark{2}Fudan University, \IEEEauthorrefmark{3}Shanghai Innovation Institute\\
\{wumy21, jschen23, bwwan25\}@m.fudan.edu.cn, \{peichen19, ghong, bcan20, xdpan, jrdai, m\_yang\}@fudan.edu.cn
}
\IEEEauthorblockA{
\Letter Co-corresponding authors
}
}

\begin{document}

\title{MCPZoo: A Large-Scale Dataset of Runnable Model Context Protocol Servers for AI Agent}

\maketitle
\thispagestyle{plain}
\pagestyle{plain} 





\begin{abstract}

Model Context Protocol (MCP) enables agents to interact with external tools, yet empirical research on MCP is hindered by the lack of large-scale, accessible datasets. We present MCPZoo, the largest and most comprehensive dataset of MCP servers collected from multiple public sources, comprising 129,059 servers (56,053 distinct). MCPZoo includes 16,356 server instances that have been deployed and verified as runnable and interactable, supporting realistic experimentation beyond static analysis. The dataset provides unified metadata and access interfaces, enabling systematic exploration and interaction without manual deployment effort. MCPZoo is released as an open and accessible resource to support research on MCP-based systems and security analysis.

\end{abstract}

\begin{IEEEkeywords}
Model Context Protocol, MCP, Agents, Agent Security, Security Measurement
\end{IEEEkeywords}


\IEEEpeerreviewmaketitle

\section{Introduction}
The software ecosystem is witnessing a paradigm shift driven by Large Language Models (LLMs) and autonomous Agents. Just as mobile applications revolutionized digital interaction over the last decade, "Tools"—interfaces that allow Agents to perceive and manipulate the world—are becoming the fundamental building blocks of the AI era. The Model Context Protocol (MCP) has rapidly emerged as the de facto standard for connecting these AI models with diverse data sources and environments.

However, conducting large-scale empirical research on this growing ecosystem presents significant challenges. While recent studies have begun to explore the security risks and robustness of MCP integrations (e.g., prompt injection vulnerabilities in tool use), these efforts are often constrained by the difficulty of setting up testing environments. Unlike mobile apps, which are packaged as standalone binaries, MCP servers are diverse software projects with complex, often conflicting dependencies. Consequently, existing research is frequently limited to small, manually curated datasets, leading to experiments that are difficult to reproduce or generalize.

Inspired by AndroZoo~\cite{androzoo}, which democratized Android research by collecting millions of apps for the community, we present MCPZoo. To the best of our knowledge, MCPZoo is the largest collection of Model Context Protocol servers to date. While AndroZoo solved the problem of access, MCPZoo aims to solve the problem of runnability. 


By providing a standardized, unrestricted, and scalable dataset of runnable agents, we aim to enable the research community to move beyond small-scale testing and conduct comprehensive, reproducible analyses on the security and capability of the AI agent ecosystem.

\noindent\textbf{Contribution.} MCPZoo makes the following contributions:
\begin{itemize}
    \item The largest and most comprehensive dataset of MCP servers currently available, with 129,059 MCP servers.
    \item The first dataset to include 16,356 runnable and interactable server instances.
    \item Released as an open and accessible dataset to support community-driven MCP analysis.
\end{itemize}

\section{Build Infrastructure}
\subsection{Construct MCP Zoo}



\noindent\textbf{Sources.}
We collected MCP server projects from eight public sources: MCP Store~\cite{mcpstore}, MCP World~\cite{mcpworld}, MCP Market~\cite{mcpmarket}, MCP Servers Repository~\cite{mcprepository}, AIbase MCP~\cite{aibasemcp}, Pulse MCP~\cite{pulsemcp}, MCP.so~\cite{mcpso}, and Smithery~\cite{smithery}.
Each source provides an independently curated directory of MCP servers, reflecting different community and platform perspectives.

\noindent\textbf{Data Fields.}
Both metadata and source code for each MCP server are saved. We extract a unified set of metadata fields, including the server name, description, overview text~(detailed Readme), source code URL, author information, etc. In parallel, we download the corresponding code repositories and organize all artifacts in a structured storage layout. To support cross-source alignment and efficient access, we generate symbolic links that associate multiple listings with their code snapshots.

\noindent\textbf{De-Duplication.}
To ensure high data quality, we de-duplicate large numbers of forks, mirrors, and copy-paste repositories. Our de-duplication process follows three steps. We first normalize repository URLs and clean metadata to resolve redirects and naming inconsistencies. We then vectorize representative textual content extracted from configuration and code files to capture semantic similarity between server implementations. Finally, we cluster highly similar servers and retain a single representative per cluster, while preserving provenance information for all sources.

\subsection{Make the Zoo Alive}




\noindent\textbf{Building Docker Image.}
To make servers runnable at scale, we employ an automated build agent to convert the MCP server source code into executable Docker images. The agent performs context-aware analysis of code and dependencies to infer runtime settings and generate Docker configurations. When build failures occur, it iteratively analyzes error logs and regenerates the Docker files with targeted adjustments until the image builds successfully or a retry limit is reached.

\noindent\textbf{Liveness Check.}
A successfully built container does not necessarily indicate a runnable MCP server. We therefore perform a standardized liveness check to validate whether a server is practically usable in the MCP setting. 
We perform a two-step check to guarantee the liveness.
First, we examine the containers' execution capability,
accounting for common failure cases such as missing configurations or hard-coded paths. 
Then we examine the server containers' protocol-level functionality through basic interactions, including retrieving declared tool lists. 
Only servers that pass both execution and interaction checks are considered alive in MCPZoo.
\section{MCPZoo}

\subsection{Overview}

MCPZoo is a large-scale, continuously growing dataset of MCP servers collected from multiple public sources. 
Up to December 25th, 2025, MCPZoo contains 129,059 MCP servers, corresponding to 56,053 distinct servers after de-duplication. Among them, 16,356 servers are verified to be runnable. The collected source code amounts to 399 GB in total. Table~\ref{tab:source-stats} summarizes the composition of MCPZoo by source, reporting the number of collected servers from each of the platforms.

\begin{table}[ht]
  \centering
  \caption{Statistics of MCP Servers from Multiple Public Sources~(Dec. 2025)}
  \label{tab:source-stats}
  \small
  \begin{tabular}{lcc}
    \toprule
    \textbf{Source} & \textbf{\# Collected} &  \textbf{Percentage} \\
    \midrule
    MCP Store        & 39,632 & 30.71\% \\
    MCP World        & 31,048 & 24.05\% \\
    MCP Market       & 16,105 & 12.48\% \\
    MCP Repository   & 14,341 & 11.11\% \\
    AIbase MCP       & 11,120 & 8.62\%  \\
    Pulse MCP        & 6,884  & 5.33\%  \\
    MCP.so           & 6,772  & 5.25\%  \\
    Smithery         & 3,157  & 2.45\%  \\
    \midrule
    \textbf{Total}   & 129,059 & 100.00\% \\
    \textbf{Total~(Distinct)}   & 56,053 & -- \\
    \bottomrule
  \end{tabular}
\end{table}

The dataset is constructed through an ongoing collection effort and is actively expanding as new MCP servers emerge in the ecosystem.

\subsection{Visit MCPZoo}

MCPZoo is publicly accessible through a website\footnote{https://security.fudan.edu.cn/zoo}. The website provides downloadable metadata covering the entire collection of MCP servers, enabling researchers to systematically inspect and filter the dataset at scale.
A summary of all available metadata fields is presented in Table~\ref{tab:metadata-fields}.

\begin{table}[ht]
    \centering
    \caption{Metadata fields provided by MCPZoo}
    \label{tab:metadata-fields}
    \small
    \begin{tabular}{p{0.22\linewidth} p{0.7\linewidth}}
        \toprule
        \textbf{Field} & \textbf{Description} \\
        \midrule
        \texttt{serverName} &
        The human-readable name of the MCP server as provided by the source platform. \\

        \texttt{description} &
        A short textual description summarizing the functionality and purpose of the server. \\

        \texttt{source} &
        The originating platform or repository from which the MCP server was collected. \\

        \texttt{creator} &
        The declared creator or organization of the MCP server. \\

        \texttt{overview} &
        A more detailed overview of the server, typically describing its supported tools, usage scenarios, or design intent. \\
        \bottomrule
    \end{tabular}
\end{table}


MCPZoo provides the capability to dynamically invoke MCP servers through remote connections. At present, 16,356 servers have been deployed, verified, and confirmed to be interactable in real execution environments. For each server, MCPZoo maintains a unified remote access configuration, which allows users to directly connect to and interact with MCP servers in a consistent manner. 
This design lowers the barrier to testing and experimentation.
The MCPZoo website publicly exposes access interfaces for 15 runnable MCP servers for trial use. Researchers who require access to a larger set of interactable servers may contact us for extended access.



\subsection{Access Conditions}


MCPZoo is made available to the research community to support measurement, systems, and security studies on the MCP ecosystem.
We require that researchers requesting access to MCPZoo agree to the following conditions.
(1) \textit{Legal Compliance.}
Researchers must assess the legal implications of downloading, storing, and analyzing the collected materials in accordance with applicable local laws and host institution policies.
(2) \textit{No Redistribution.}
The dataset should not, in general, be redistributed or republished, either in full or in part.
(3) \textit{Non-Commercial Use.}
The dataset must not be used for commercial purposes.
(4) \textit{Responsible Interaction.}
Researchers are expected to interact with MCP servers in a responsible manner and to avoid disruptive, abusive, or harmful behaviors toward deployed services. Security testing beyond benign interaction or measurement should follow responsible disclosure principles and may require additional approval.
(5) \textit{Faculty Endorsement.}
Access requests must be endorsed by a faculty member or an individual in a permanent research position, who agrees to these conditions and commits to responsible use of the dataset.

We \textbf{kindly request} that the use of MCPZoo be acknowledged, and \textbf{encourage} researchers to disclose the list of the MCP servers involved in their studies whenever feasible, to support reproducibility.
\section{Leveraging MCPZoo}



MCPZoo can be used as a shared experimental foundation for research on real-world MCP deployments. 

First, MCPZoo can enable principled agent benchmarking by providing diverse and realistic tool environments for evaluating agents’ capabilities.
Benchmarking plays a critical role in emerging research areas~\cite{yang2025mcpsecbench,lian2025ase}. By supporting repeatable interaction with real MCP servers, MCPZoo makes it possible to construct authoritative, protocol-level benchmarks for MCP-based systems, complementing recent efforts that establish systematic evaluation frameworks for MCP security and AI-generated code at scale.

Second, MCPZoo can support large-scale security analysis from both attack and defense perspectives.
Recent studies~\cite{guo2025systematicmcp,wang2025mcpguard} and industrial tools~\cite{mcp_scan2025,Tencent_AI-Infra-Guard_2025} have begun to identify various attack vectors against MCP systems, e.g., prompt injection, sandbox escape, and tool misuse. A large-scale, real-world dataset such as MCPZoo enables systematic examination of these vulnerabilities across diverse MCP implementations, moving beyond isolated case studies. At the same time, the availability of runnable servers provides a practical foundation for studying and validating defense mechanisms under realistic execution settings.

Finally, MCPZoo can facilitate ecosystem- and protocol-level evolution studies.
For example, \citet{hou2025mcpreview} systematically characterized security threats across the lifecycle of MCP.
\citet{guo2025mcpecosystem} conducted a large-scale empirical study of the MCP ecosystem from the perspectives of markets, servers, and clients, providing evidence-based insights into ecosystem-level security risks.
By enabling empirical analysis of MCP adoption, implementation patterns, and cross-project diversity across a large collection of real-world MCP servers, MCPZoo offers a unique opportunity to study how the MCP ecosystem evolves in practice and to inform the future design of the protocol.
\section{Conclusion}

We present MCPZoo, a large-scale dataset of MCP servers collected from multiple public sources, with a large number of servers verified as runnable and interactable. The dataset is publicly available, together with access interfaces, to support future research on MCP-based systems and analysis. MCPZoo can be accessed through the website: https://security.fudan.edu.cn/zoo.


\bibliographystyle{IEEEtranN}
\bibliography{reference}


\end{document}